\documentstyle[sprocl,epsf,rotate]{article}

\def\be{\begin{equation}}
\def\ee{\end{equation}}
\def\bea{\begin{eqnarray}}
\def\eea{\end{eqnarray}}
\def\lesssim{\mathrel{\hbox{\rlap{\hbox{\lower4pt\hbox{$\sim$}}}\hbox{$<$}}}}
\def\gtrsim{\mathrel{\hbox{\rlap{\hbox{\lower4pt\hbox{$\sim$}}}\hbox{$>$}}}}

\begin{document}

%%% start CERN preprint title page %%%%%%%%%%%%%
\begin{titlepage}
 
\begin{flushright}
CERN-TH/97-241\\
hep-ph/9709291
\end{flushright}
 
\vspace{1.5cm}
 
\begin{center}
\Large\bf CP Violation Beyond the Standard Model
\end{center}
 
\vspace{1.2cm}
 
\begin{center}
Robert Fleischer\\
{\sl Theory Division, CERN, CH-1211 Geneva 23, Switzerland}
\end{center}
 
\vspace{1.3cm}

\begin{center}
{\bf Abstract}\\[0.3cm]
\parbox{11cm}{
Recent developments concerning CP violation beyond the Standard Model are 
reviewed. The central target of this presentation is the $B$ system, as it 
plays an outstanding role in the extraction of CKM phases. Besides a general 
discussion of the appearance of new physics in the corresponding CP-violating 
asymmetries through $B^0_q$--$\overline{B^0_q}$ mixing $(q\in\{d,s\})$, it 
is emphasized that CP violation in non-leptonic penguin modes, e.g.\ in 
$B_d\to\phi\, K_{\rm S}$, offers a powerful tool to probe physics beyond 
the Standard Model. In this respect $B\to\pi K$ modes, which have been 
observed recently by the CLEO collaboration, may also turn out to be very 
useful. Their combined branching ratios allow us to constrain the CKM angle 
$\gamma$ and may indicate the presence of physics beyond the Standard Model.}
\end{center}
 
\vspace{1cm}
 
\begin{center}
{\sl To appear in the Proceedings of the\\
7th International Symposium on Heavy Flavor Physics\\
Santa Barbara, California, 7--11 July 1997}
\end{center}
 
\vspace{1.5cm}
 
\vfil
\noindent
CERN-TH/97-241\\
September 1997
 
\end{titlepage}
 
\thispagestyle{empty}
\vbox{}
\newpage
 
\setcounter{page}{1}
 
%%% end CERN preprint title page %%%%%%%%%%%%%

\title{CP VIOLATION BEYOND THE STANDARD MODEL}

\author{ ROBERT FLEISCHER }

\address{Theory Division, CERN\\ 
CH-1211 Geneva 23, Switzerland}


\maketitle\abstracts{
Recent developments concerning CP violation beyond the Standard Model are 
reviewed. The central target of this presentation is the $B$ system, as it 
plays an outstanding role in the extraction of CKM phases. Besides a general 
discussion of the appearance of new physics in the corresponding CP-violating 
asymmetries through $B^0_q$--$\overline{B^0_q}$ mixing $(q\in\{d,s\})$, it 
is emphasized that CP violation in non-leptonic penguin modes, e.g.\ in 
$B_d\to\phi\, K_{\rm S}$, offers a powerful tool to probe physics beyond 
the Standard Model. In this respect $B\to\pi K$ modes, which have been 
observed recently by the CLEO collaboration, may also turn out to be very 
useful. Their combined branching ratios allow us to constrain the CKM angle 
$\gamma$ and may indicate the presence of physics beyond the Standard Model.}
  
\section{Introduction}

CP violation is one of the least well understood and least experimentally 
tested phenomena in present particle physics. There are several reasons to 
expect physics beyond the Standard Model to show up in CP-violating effects.
First, the Standard Model description of CP violation, i.e.\ the phase 
structure of the Cabibbo-Kobayashi-Maskawa matrix (CKM matrix)~\cite{ckm}, 
has not yet been tested experimentally. Second, many extensions of the Standard
Model have additional sources of CP violation or may affect Standard Model 
relations among CP-violating observables~\cite{new-phys}. Third, CP violation 
is one of the three necessary conditions for the large imbalance between 
matter and antimatter that is observed in the Universe~\cite{sakharov}.
Calculations within the framework of the Standard Model show, however, that 
CP violation seems to be too small to generate this imbalance. This feature 
could be a hint for the need of sources for CP violation beyond the Standard 
Model.

At present the observed ``indirect'' CP violation in the neutral $K$-meson 
system~\cite{vel} can successfully be described by the Standard Model. Since 
so far only a single CP-violating observable, $\varepsilon$, has to be fitted,
it is, however, not surprising that many different ``non-standard'' model 
descriptions of CP \mbox{violation} are imaginable~\cite{new-phys}. While a 
measurement of Re$(\varepsilon'/\varepsilon)\not=0$ describing ``direct'' 
CP violation in the neutral kaon system would exclude ``superweak'' scenarios 
of CP violation~\cite{superweak} in an unambiguous way, this observable will 
not allow a stringent test of the Standard Model description of CP violation, 
unless the presently large theoretical uncertainties related to poorly known 
hadronic matrix elements can be controlled in a reliable way~\cite{bf-rev}. 
More promising with respect to testing the CP-violating sector of the Standard 
Model are the rare decays $K_{\rm L}\to\pi^0\,\overline{\nu}\,\nu$ and 
$K^+\to\pi^+\,\overline{\nu}\,\nu$, as has been stressed by Andrzej Buras at 
this symposium~\cite{buras}. 

It is obvious from the brief discussion given above that the $K$-meson 
system by itself cannot provide the whole picture of CP violation. Therefore 
it is essential to study CP violation outside this system. In this respect, 
the $B$ system appears to be most promising, which is also reflected by the 
tremendous experimental efforts at future $B$ factory facilities~\cite{B-exp}.
Let me note that there are also other interesting systems to explore 
CP violation and to search for physics beyond the Standard Model, e.g.\ 
the $D$-meson system~\cite{D-rev}, where sizeable mixing or CP-violating 
effects would signal new physics because of the tiny Standard Model 
``background''. In this presentation I unfortunately cannot discuss these 
systems in more detail and shall focus on $B$ decays.

\section{The Central Target: CP Violation in the B System}

\subsection{General Remarks}

It is by now well-known that large CP-violating effects are expected to show 
up in non-leptonic $B$-meson decays~\cite{cp-revs,rev}. In several cases the 
corresponding CP-violating observables are closely related to the angles 
$\alpha$, $\beta$ and $\gamma$ of the usual ``non-squashed'' unitarity 
triangle~\cite{ut} of the CKM matrix. Probably the most prominent example is 
the ``gold-plated'' mode $B_d\to J/\psi\,K_{\rm S}$, measuring $\sin(2\beta)$ 
in a clean way through ``mixing-induced'' CP violation~\cite{csbs}. 
Another ``benchmark'' mode is $B_d\to\pi^+\pi^-$, which at first sight 
seems to measure $\sin(2\alpha)$. A closer look shows, however, that in 
contrast to $B_d\to J/\psi\,K_{\rm S}$ penguin contributions may lead to 
serious hadronic uncertainties in the former decay requiring more involved 
strategies to extract a reliable value of $\alpha$~\cite{cp-revs,rev}. 
Although theoretical clean techniques to determine the third angle $\gamma$ 
of the unitarity triangle are on the market, using for instance the ``tree'' 
decays $B^\pm\to DK^\pm$ or $B_s\to D_s K$ (see e.g.\ Ref.~\cite{rev} for a 
recent review), these methods are in general very challenging from an 
experimental point of view. It should also be kept in mind that a generic 
problem of the determination of the angles of the unitarity triangle from 
CP-violating observables is that one has to deal with discrete 
ambiguities~\cite{ambig}.

The presence of new physics could manifest itself in several ways, e.g.\
through a violation of the unitarity relation 
\begin{equation}\label{ut-rel}
\alpha+\beta+\gamma=180^\circ.
\end{equation}
Another possibility is to find that Eq.~(\ref{ut-rel}) is satisfied, but
that the directly measured angles disagree with the Standard Model 
expectation, in particular with measurements of the sides of the unitarity
triangle through semileptonic $b\to c\,l\,\overline{\nu}_l$, $b\to u\,l\,
\overline{\nu}_l$ decays and $B^0_d$--$\overline{B^0_d}$ mixing~\cite{bf-rev}.

The goal of future $B$ physics experiments~\cite{B-exp} is therefore to
perform as many independent CP-violating measurements as possible to 
overconstrain the unitarity triangle as much as possible. Either these 
measurements will lead to results that are consistent with each other and 
with the Standard Model expectations, leading eventually to the full 
determination of the CKM matrix, or discrepancies may show up that could 
shed light on new physics. Needless to note, the latter option would be 
much more exciting.  

\subsection{Theoretical Ingredients}

In order to analyse the impact of physics beyond the Standard Model, we have 
to briefly recapitulate the theoretical ingredients that are at the basis of 
direct measurements of CKM phases through CP-violating asymmetries, as
sketched above. The central role is played by non-leptonic $B$ decays into
final CP eigenstates $|f\rangle$. In that case the corresponding 
time-dependent CP asymmetries can be expressed as
\begin{eqnarray}
\lefteqn{a_{\mbox{{\scriptsize CP}}}(t)\equiv\frac{\Gamma(B^0_q(t)\to f)-
\Gamma(\overline{B^0_q}(t)\to f)}{\Gamma(B^0_q(t)\to f)+
\Gamma(\overline{B^0_q}(t)\to f)}=}\nonumber\\
&&{\cal A}^{\mbox{{\scriptsize dir}}}_{\mbox{{\scriptsize CP}}}(B_q\to f)
\cos(\Delta M_q\,t)+{\cal A}^{\mbox{{\scriptsize
mix--ind}}}_{\mbox{{\scriptsize CP}}}(B_q\to f)\sin(\Delta M_q\,t)
\,.\label{ee6}
\end{eqnarray}
Here direct CP violation has been separated from mixing-induced CP violation
by introducing
\begin{equation}\label{ee7}
{\cal A}^{\mbox{{\scriptsize dir}}}_{\mbox{{\scriptsize CP}}}(B_q\to f)\equiv
\frac{1-|\xi_f^{(q)}|^2}{1+|\xi_f^{(q)}|^2}\,,\quad
{\cal A}^{\mbox{{\scriptsize mix--ind}}}_{\mbox{{\scriptsize
CP}}}(B_q\to f)\equiv\frac{2\,\mbox{Im}\,\xi^{(q)}_f}{1+
|\xi^{(q)}_f|^2}\,.
\end{equation}
The observable $\xi^{(q)}_f$ contains essentially all the information that 
is needed to evaluate these asymmetries and will be discussed in more detail 
below. 

In most decays $B_q\to f$ of interest for extracting CKM phases, only 
mixing-induced CP violation shows up. There are two conditions for a clean 
relation of ${\cal A}^{\mbox{{\scriptsize mix--ind}}}_{\mbox{{\scriptsize 
CP}}}(B_q\to f)$ to angles of the unitarity triangle:
\begin{itemize}
\item The ratio $|\Gamma_{12}^{(q)}|/|M_{12}^{(q)}|$ of the off-diagonal 
elements of the decay and mass matrices describing $B^0_q$--$\overline{B^0_q}$
mixing has to be much smaller than 1. 
\item The decay amplitude $A(B_q\to f)$ has to be dominated by
a single weak amplitude implying vanishing direct CP violation, i.e.\
${\cal A}^{\mbox{{\scriptsize dir}}}_{\mbox{{\scriptsize CP}}}(B_q\to f)=0$.
\end{itemize}
If these two requirements are met simultaneously, the observable $\xi^{(q)}_f$
is given by a pure phase factor, i.e.\ $|\xi^{(q)}_f|=1$, and can be 
expressed as~\cite{ns}
\begin{equation}\label{xi-exp}
\xi^{(q)}_f=\eta_{\mbox{{\scriptsize CP}}}^f\cdot\left(\frac{X_f}{X_f^\ast}
\right)\cdot\left(\frac{Y_q}{Y_q^\ast}\right)\cdot\left(\frac{Z_f}{Z_f^\ast}
\right)\,,
\end{equation}
where $X_f/X_f^\ast$ is the weak decay phase, $Y_q/Y_q^\ast$ denotes the weak
$B^0_q$--$\overline{B^0_q}$ mixing phase, and the last factor $Z_f/Z_f^\ast$ is
only needed if the CP eigenstate $|f\rangle$ satisfying $({\cal CP})|f\rangle
=\eta_{\mbox{{\scriptsize CP}}}^f|f\rangle$ contains a neutral $K$ meson as
is, for instance, the case in $B_d\to J/\psi\, K_{\rm S}$. 

\subsection{What about New Physics?}

There are several excellent reviews~\cite{new-phys} dealing with this 
question, where much more detailed discussions can be found. Here I have to 
be rather general. 

Let me first note that it is very difficult to change $|\Gamma_{12}^{(q)}|/
|M_{12}^{(q)}|\ll1$ through new physics. To this end one would need a new 
dominant contribution to tree decays of $B_q$ mesons, which is very unlikely, 
or a strong suppression of the mixing compared to the Standard Model, which 
is also unlikely but not impossible for the $B_s$ system. 

The $Z_f/Z_f^\ast$ factor in Eq.~(\ref{xi-exp}) may be different from the 
Standard Model value if new physics shows up in $K^0$--$\overline{K^0}$ 
mixing. An interesting test is provided by the relations~\cite{ns}
\begin{eqnarray}
{\cal A}^{\mbox{{\scriptsize mix--ind}}}_{\mbox{{\scriptsize
CP}}}(B_d\to D^+D^-)&=&-\,{\cal A}^{\mbox{{\scriptsize 
mix--ind}}}_{\mbox{{\scriptsize CP}}}(B_d\to J/\psi\,K_{\rm S})\,\,\,=\,\,\,
\sin(2\beta)\\
{\cal A}^{\mbox{{\scriptsize mix--ind}}}_{\mbox{{\scriptsize
CP}}}(B_s\to D_s^+D_s^-)&=&-\,{\cal A}^{\mbox{{\scriptsize 
mix--ind}}}_{\mbox{{\scriptsize CP}}}(B_s\to J/\psi\,K_{\rm S})\,\,\,\approx
\,\,\,0
\end{eqnarray}
holding within the Standard Model. Although $|Z_f|$ may be affected 
by new physics, arg$(Z_f)$ can only be changed in very contrived models. This 
feature is guaranteed by the small value of $\varepsilon$ parametrizing 
indirect CP violation in the kaon system~\cite{ns}.

Typically new physics is expected to show up at a scale $\Lambda$ in the
TeV regime so that its effects in $W$-mediated, CKM-allowed tree-level
processes are highly suppressed by ${\cal O}(M_W^2/\Lambda^2)$. However,
owing to the loop-suppression of ``rare'' flavour-changing neutral current 
(FCNC) processes, it is plausible that new physics contributions could
there be of similar magnitude as those of the Standard Model. Consequently 
one expects sizeable effects either in $B^0_q$--$\overline{B^0_q}$ mixing 
affecting the phase factor $Y_q/Y_q^\ast$, or in the amplitudes of 
penguin-dominated decays, e.g.\ $B_d\to\phi\,K_{\rm S}$, affecting the 
$X_f/X_f^\ast$ phase factor in Eq.~(\ref{xi-exp}). Specific examples 
for new physics are models with four generations, extended Higgs sectors, 
$Z$-mediated FCNCs, non-minimal supersymmetric models, and many others 
discussed extensively in several reviews about that topic~\cite{new-phys}. 
Before turning to penguin modes, let me first discuss the appearance of 
new physics through $B^0_q$--$\overline{B^0_q}$ mixing in more detail. 

\section{The Manifestation of New Physics through $B^0_q$--$\overline{B^0_q}$ 
Mixing}
 
Since $B^0_q$--$\overline{B^0_q}$ mixing originating from the well-known box 
diagrams is already a one-loop effect in the Standard Model, it is plausible 
that new physics may affect this phenomenon considerably~\cite{ns}. The
corresponding mixing-induced CP-violating asymmetries are affected in 
particular by a possible shift of the weak $B^0_q$--$\overline{B^0_q}$ mixing 
phases from their Standard Model values through new physics:
\begin{eqnarray}
\phi_{\rm M}^{(d)}&=&2\beta+2\phi_{\rm new}^{(d)}\\
\phi_{\rm M}^{(s)}&=&0+2\phi_{\rm new}^{(s)}\,.
\end{eqnarray}
As far as the ``benchmark'' modes $B_d\to J/\psi\,K_{\rm S}$ and 
$B_d\to\pi^+\pi^-$ are concerned, their mixing-induced CP asymmetries are 
modified as follows (note that penguin contributions are neglected in 
$B_d\to\pi^+\pi^-$):
\begin{eqnarray}
{\cal A}^{\mbox{{\scriptsize mix--ind}}}_{\mbox{{\scriptsize
CP}}}(B_d\to J/\psi\,K_{\rm S})&=&-\,\sin(2\beta+2\phi_{\rm new}^{(d)})
\,\,\equiv\,\,-\,\sin(2\beta_{\rm exp})\\
{\cal A}^{\mbox{{\scriptsize mix--ind}}}_{\mbox{{\scriptsize
CP}}}(B_d\to\pi^+\pi^-)&=&-\,\sin(2\alpha-2\phi_{\rm new}^{(d)})
\,\,\equiv\,\,-\,\sin(2\alpha_{\rm exp})\,,
\end{eqnarray}
so that these observables do not probe the angles $\alpha$ and $\beta$ of
the unitarity triangle but
\begin{equation}
\alpha_{\rm exp}\equiv\alpha-\phi^{(d)}_{\rm new}\,,\quad
\beta_{\rm exp}\equiv\beta+\phi^{(d)}_{\rm new}\,.
\end{equation}
In the sum of these experimentally determined angles the new physics phase
$\phi^{(d)}_{\rm new}$ cancels, however, so that we have
\begin{equation}
\alpha_{\rm exp}+\beta_{\rm exp}=\alpha+\beta\,,
\end{equation}
as was pointed out by Nir and Silverman~\cite{ns}. Consequently, in order to 
test the unitarity relation Eq.~(\ref{ut-rel}) thereby searching for physics 
beyond the Standard Model, it is crucial to determine $\gamma$ in a variety
of ways. The ``standard'' strategies to extract this angle~\cite{rev} can be 
divided into two categories that are unfortunately both difficult to perform 
in practice as I have already noted. The first one uses charged 
$B^\pm\to D K^\pm$ and related decays~\cite{BDK}, which are pure tree modes 
where no FCNCs are involved. Therefore new physics should play a minor 
role in these channels and one expects to find
\begin{equation}
\alpha_{\rm exp}+\beta_{\rm exp}+\gamma_{\rm exp}^{(1)}=180^\circ,
\end{equation}
where $\gamma_{\rm exp}^{(1)}$ denotes the value of $\gamma$ that is 
determined from these modes. In the second category~\cite{adk}, one employs 
decays such as $B_s\to D_sK$, which are also pure tree modes, and determines 
an experimental value $\gamma_{\rm exp}^{(2)}$ for $\gamma$ with the help of 
$B^0_s$--$\overline{B^0_s}$ mixing. Since here a loop-induced FCNC process -- 
the mixing -- and the related CP-violating weak phase enter, it is well 
possible to find
\begin{equation}
\alpha_{\rm exp}+\beta_{\rm exp}+\gamma_{\rm exp}^{(2)}\not=180^\circ
\end{equation}
because of $\phi_{\rm new}^{(s)}$. The presence of this phase would also be
signalled by sizeable CP-violating effects in $B_s\to J/\psi\,\phi$ or
$B_s\to D_s^+D_s^-$ exhibiting tiny CP violation within the Standard Model
due to the small $B_s^0$--$\overline{B^0_s}$ mixing phase~\cite{ns}.

The lesson that we have learned from these considerations is that it does 
not suffice to measure only $\alpha_{\rm exp}$ and $\beta_{\rm exp}$. It is 
essential to determine the third angle $\gamma$ in several ways, which is
unfortunately an experimental challenge.
  
As we have just seen, new physics can affect CP violation in the ``benchmark''
modes to extract CKM phases mainly through contributions to 
$B^0_q$--$\overline{B^0_q}$ mixing. The same new physics is, however, expected 
to manifest itself also in other FCNC processes, e.g.\ in $b\to s$ penguin 
modes and other rare decays. This is in fact the case in specific model 
calculations (see e.g.\ Ref.~\cite{model-calcs}). \mbox{Models} of new 
physics can in principle be distinguished by their contributions to such 
processes \cite{disting}, and in order to get ``the whole picture'', it is 
important to measure both CP asymmetries and rare decays. The effects of new 
physics in the ``usual'' rare $B$ decays~\cite{ali} have been reviewed by 
JoAnne Hewett at this symposium~\cite{hew}. In the subsequent section I would 
like to turn to another class of rare decays, the penguin-induced non-leptonic 
$B$ decays.

\section{CP Violation in Penguin Modes as a Probe of New Physics}

Let me begin the discussion of these decays by focusing on the $b\to d$ 
penguin mode $B_d\to K^0\overline{K^0}$. An analysis of new-physics effects
in this channel was performed, e.g.\ in Ref.~\cite{mpw}. If one assumes that 
penguins with internal top quarks play the dominant role in this transition, 
the weak $B^0_d$--$\overline{B^0_d}$ mixing and $B_d\to K^0\overline{K^0}$ 
decay phases cancel each other in the corresponding observable 
$\xi^{(d)}_{K^0\overline{K^0}}$, implying vanishing CP violation in 
that decay. Consequently one would conclude that a measurement of 
non-vanishing CP violation in $B_d\to K^0\overline{K^0}$ would signal physics 
beyond the Standard Model. However, long-distance effects related to penguins 
with internal charm and up quarks may easily spoil the assumption of 
top-quark dominance~\cite{rev,bf1}. As was pointed out in 
Ref.~\cite{my-KK.bar}, these contributions may lead to sizeable CP violation 
in $B_d\to K^0\overline{K^0}$ even within the Standard Model, so that a 
measurement of such CP asymmetries would not necessarily imply new physics, 
as claimed in several previous papers. Unfortunately a measurement of these 
effects will be very difficult since the Standard Model expectation for the 
corresponding branching ratio is ${\cal O}(10^{-6})$ which is still one order 
of magnitude below the recent CLEO bound~\cite{cleo} $\mbox{BR}(B_d\to 
K^0\overline{K^0})<1.7\cdot10^{-5}$.

More promising in this respect and -- more importantly -- to search for 
physics beyond the Standard Model is the $b\to s$ penguin mode $B_d\to\phi\, 
K_{\rm S}$. The branching ratio for this decay is expected to be of 
${\cal O}(10^{-5})$ and may be large enough to investigate this channel 
at future $B$ factories. Interestingly there is, to a very good 
approximation, no non-trivial CKM phase present in the corresponding decay 
amplitude~\cite{rev}, so that direct CP violation vanishes and mixing-induced 
CP violation measures simply the weak $B^0_d$--$\overline{B^0_d}$ mixing 
phase. It should be stressed that this statement does {\it not} require the 
questionable assumption of top-quark dominance in penguin amplitudes. 
Consequently an important probe for new physics in $b\to s$ FCNC processes 
is provided by the relation
\begin{equation}
{\cal A}^{\mbox{{\scriptsize mix--ind}}}_{\mbox{{\scriptsize
CP}}}(B_d\to J/\psi\, K_{\mbox{{\scriptsize S}}})={\cal 
A}^{\mbox{{\scriptsize mix--ind}}}_{\mbox{{\scriptsize
CP}}}(B_d\to \phi\, K_{\mbox{{\scriptsize S}}})
\end{equation}
holding within the Standard Model framework. The theoretical accuracy 
of this relation is limited by certain neglected terms that are 
CKM-suppressed by ${\cal O}(\lambda^2)$ and may lead to tiny direct 
CP-violating asymmetries in $B_d\to\phi\, K_{\rm S}$ of at most 
${\cal O}(1\%)$ \cite{rev}. Recently the importance of $B_d\to\phi\, 
K_{\rm S}$ and similar modes such as $B_d\to\eta' K_{\rm S}$ to search 
for new physics in $b\to s$ transitions has been emphasized by several 
authors~\cite{rev,model-calcs,loso}. It is possible that new physics 
affects both $b\to s$ penguin decays and $B^0_s$--$\overline{B^0_s}$ mixing. 
While $B^0_s$--$\overline{B^0_s}$ mixing is difficult to measure because of
the large mixing parameter, the penguin modes appear to be more 
promising from an experimental point of view. 

\section{Searching for $\gamma$ and New Physics with $B\to\pi K$ Modes}

A simple approach to determine $\gamma$ with the help of the branching ratios 
for $B^+\to\pi^+K^0$, $B^0_d\to\pi^-K^+$ and their charge conjugates was 
proposed in Ref.~\cite{PAPIII} (see also Ref.~\cite{rev}). It makes use of 
the fact that the general phase structure of the corresponding decay 
amplitudes is known reliably within the Standard Model. Moreover it employs 
the $SU(2)$ isospin symmetry of strong interactions to relate the QCD penguin 
contributions. If the magnitude of the current-current amplitude $T'$ 
contributing to $B^0_d\to\pi^-K^+$ is known -- it can be fixed e.g.\ through 
$B^+\to\pi^+\pi^0$, ``factorization'', or hopefully lattice gauge theory 
one day -- two amplitude triangles can be constructed, allowing in particular 
the extraction of $\gamma$. This approach is promising for future 
$B$-physics experiments since it requires only time-independent measurements 
of branching ratios at the ${\cal O}(10^{-5})$ level. If one measures in 
addition the branching ratios for $B^+\to\pi^0K^+$ and its charge-conjugate, 
also the $b\to s$ electroweak penguin amplitude can be determined, which is 
another interesting probe for new physics~\cite{PAPI}.

Recently the CLEO collaboration~\cite{cleo} has reported the first observation
of the decays $B^+\to\pi^+K^0$ and $B^0_d\to\pi^-K^+$. At present, however, 
only combined branching ratios, i.e.\ averaged ones over decays and their 
charge-conjugates, are available with large experimental uncertainties. 
Therefore it is not yet possible to extract $\gamma$ from the triangle 
construction proposed in Ref.~\cite{PAPIII}. The recent CLEO measurements 
allow, however, to derive interesting {\it constraints} on $\gamma$, which 
are of the form 
\begin{equation}\label{gamma-bound}
0^\circ\leq\gamma\leq\gamma_0\quad\lor\quad180^\circ-
\gamma_0\leq\gamma\leq180^\circ 
\end{equation}
and are hence complementary to the presently allowed range of
\begin{equation}\label{UT-fits}
42^\circ\lesssim\gamma\lesssim135^\circ
\end{equation}
for that angle arising from the usual fits of the unitarity triangle 
\cite{bf-rev}. This remarkable feature has been pointed out recently by
Mannel and myself in Ref.~\cite{fm2}. The quantity $\gamma_0$ in Eq.\ 
(\ref{gamma-bound}) depends both on the ratio 
\begin{equation}
R\equiv\frac{\mbox{BR}(B_d\to\pi^\mp K^\pm)}{\mbox{BR}(B^\pm\to\pi^\pm K)} 
=\frac{\mbox{BR}(B_d^0\to\pi^- K^+)+\mbox{BR}(\overline{B_d^0}\to\pi^+ 
K^-)}{\mbox{BR}(B^+\to\pi^+K^0)+\mbox{BR}(B^-\to\pi^-\overline{K^0})}
\end{equation}
of the combined branching ratios and on the amplitude ratio $r\equiv|T'|/|P'|$ 
of the current-current and penguin operator contributions to $B_d\to\pi^\mp 
K^\pm$. 

A very important special case is $R=1$. For $R>1$, the constraints on 
$\gamma$ require some knowledge about $r$. On the other hand, if $R$ is found 
experimentally to be smaller than 1, bounds on $\gamma$ can always be 
obtained independently of $r$. The point is that $\gamma_0$ takes a maximal 
value 
\begin{equation}
\gamma_0^{\rm max}=\mbox{arccos}(\sqrt{1-R})\,,
\end{equation}
depending only on the ratio $R$ of combined $B\to\pi K$ branching 
ratios~\cite{fm2}. 

Let us take as an example the central value 0.65 of the recent CLEO
measurements \cite{cleo} yielding $R=0.65\pm0.40$. This value corresponds to 
$\gamma_0^{\rm max}=54^\circ$ and implies the range $0^\circ\leq\gamma\leq
54^\circ$ $\lor$ $126^\circ\leq\gamma\leq180^\circ$, which has only the small
overlap $42^\circ\lesssim\gamma\leq54^\circ$ $\lor$ $126^\circ\leq\gamma
\lesssim135^\circ$ with the range (\ref{UT-fits}). The two pieces of this
range are distinguished by the sign of the quantity $\cos\delta$, where 
$\delta$ is the CP-conserving strong phase shift between the $T'$ and $P'$ 
amplitudes. Using arguments based on ``factorization'', one expects 
$\cos\delta>0$ corresponding to the former interval of that range, 
i.e.\ $42^\circ\lesssim\gamma\leq54^\circ$ in our example \cite{fm2} 
(see Ref.~\cite{ag} for a recent model calculation). Consequently, once 
more data come in confirming $R<1$, the decays $B_d\to\pi^\mp K^\pm$ 
and $B^\pm\to\pi^\pm K$ may put the Standard Model to a decisive test and 
could open a window to new physics. Effects of physics beyond the Standard 
Model in $B\to\pi K$ modes have been analysed in a recent paper~\cite{fm3}. 
%A striking new-physics effect would be, for instance, to
%find sizeable CP violation in the decay $B^+\to\pi^+K^0$. 
A detailed study of various implications of the bounds on $\gamma$ discussed 
above has been performed very recently in Ref.~\cite{gnps}, where the 
issue of new physics has also been addressed.

\section{Summary and Outlook}

In conclusion, we have seen that the kaon system -- the only one where CP 
violation has been observed to date -- cannot provide the whole picture of 
that phenomenon. In addition to other interesting systems, e.g. the $D$ 
system, non-leptonic $B$-meson decays, where large CP asymmetries are 
expected within the Standard Model, are extremely promising to test 
the CKM picture of CP violation. Physics beyond the Standard Model
may show up in CP-violating asymmetries of $B$ decays through several 
mechanisms. Probably the most important ones are new-physics contributions 
to $B^0_q$--$\overline{B^0_q}$ mixing, and contributions to loop-suppressed, 
``rare'' penguin-induced $B$ decays. A sensitive probe of new physics is
also provided by $B\to\pi K$ decays that have recently been observed by the 
CLEO collaboration and allow us to constrain the CKM angle $\gamma$. In 
the foreseeable future, dedicated $B$-physics experiments may bring unexpected 
results that could guide us to physics beyond the Standard Model. Certainly
a very exciting era of particle physics is ahead of us!

\section*{Acknowledgements}
I would like to thank Jeff Richman and Mike Witherell for inviting me to 
that stimulating symposium in a most enjoyable environment and for providing 
generous travel support.


\section*{References}
\begin{thebibliography}{99}
\bibitem{ckm}N. Cabibbo, {\it Phys.\ Rev.\ Lett.}~{\bf 10}, 531 (1963); 
M. Kobayashi and T.~Maskawa, {\it Progr.\ Theor.\ Phys.}~{\bf 49}, 652 (1973).
\bibitem{new-phys}For reviews see e.g.\ Y. Grossman, Y. Nir and R.
Rattazzi, hep-ph/9701231; M.~Gronau and D. London, {\it Phys.\ Rev.}~{\bf 
D55}, 2845 (1997); Y. Nir and H.R. Quinn, {\it Annu.\ Rev.\ Nucl.\ Part.\ 
Sci.}~{\bf 42}, 211 (1992).
\bibitem{sakharov}A. Sakharov, {\it JETP Lett.}~{\bf 5}, 24 (1967).
\bibitem{vel}M. Velasco, these proceedings.
\bibitem{superweak}L. Wolfenstein, {\it Phys.\ Rev.\ Lett.}~{\bf 13}, 562
(1964).
\bibitem{bf-rev}For a review see A.J. Buras and R. Fleischer, hep-ph/9704376.
\bibitem{buras}A.J. Buras, these proceedings. 
\bibitem{B-exp}P. Oddone, these proceedings; P. Karchin, these proceedings.  
\bibitem{D-rev}For a recent review see e.g.\ E. Golowich, hep-ph/9706548. 
\bibitem{cp-revs}For reviews see e.g.\ Y. Nir and H.R. Quinn, in {\it $B$ 
Decays}, ed.\ S. Stone (World Scientific, Singapore, 1994), p.\ 362; 
I. Dunietz, ibid., p.\ 393; H.R.~Quinn, {\it Nucl.\ Phys.}~{\bf B}
(Proc.\ Suppl.) {\bf 50}, 17 (1996); M. Gronau, hep-ph/9611255.
\bibitem{rev}R. Fleischer, {\it Int.\ J. Mod.\ Phys.}~{\bf A12}, 2459 (1997).
\bibitem{ut}L.L. Chau and W.-Y. Keung, {\it Phys.\ Rev.\ Lett.}~{\bf 53}, 1802
(1984); C. Jarlskog and R. Stora, {\it Phys.\ Lett.}~{\bf B208}, 268 (1988).
\bibitem{csbs}A.B. Carter and A.I. Sanda, {\it Phys.\ Rev.\ Lett.}~{\bf 45},
952 (1980); {\it Phys.\ Rev.}~{\bf D23}, 1567 (1981); I.I. Bigi and A.I.
Sanda, {\it Nucl.\ Phys.}~{\bf B193}, 85 (1981).
\bibitem{ambig}Y. Grossman, Y. Nir and M.P. Worah, hep-ph/9704287; 
Y. Grossman and H.R. Quinn, hep-ph/9705356.
\bibitem{ns}Y. Nir and D. Silverman, {\it Nucl.\ Phys.}~{\bf B345}, 301 (1990).
\bibitem{BDK}M. Gronau and D. Wyler, {\it Phys.\ Lett.}~{\bf B265}, 172 (1991);
D. Atwood, I.~Dunietz and A. Soni, {\it Phys.\ Rev.\ Lett.}~{\bf 78}, 3257 
(1997).
\bibitem{adk}R. Aleksan, I. Dunietz and B. Kayser, {\it Z. Phys.}~{\bf C54},
653 (1992); R.~\mbox{Fleischer} and I. Dunietz, 
{\it Phys.\ Lett.}~{\bf B387}, 361
(1996).
\bibitem{model-calcs}Y. Grossman and M.P. Worah, {\it Phys.\ Lett.}~{\bf B395},
241 (1997); M. Ciuchini et al., {\it Phys.\ Rev.\ Lett.}~{\bf 79}, 978 
(1997); R. Barbieri and A. Strumia, hep-ph/9704402.
\bibitem{disting}M. Gronau and D. London, Ref.~\cite{new-phys};
Y. Grossman and M. Worah, Ref.~\cite{model-calcs}.
\bibitem{ali}A. Ali, these proceedings.
\bibitem{hew}J. Hewett, these proceedings.
\bibitem{mpw}M.P. Worah, {\it Phys.\ Rev.}~{\bf D54}, 2198 (1996).
\bibitem{bf1}A.J. Buras and R. Fleischer, {\it Phys.\ Lett.}~{\bf B341}, 379 
(1995). Recently, this point has also been raised by M. Ciuchini et al.,
hep-ph/9703353.
\bibitem{my-KK.bar}R. Fleischer, {\it Phys.\ Lett.}~{\bf B341}, 205 (1994).
\bibitem{cleo}J.P. Alexander, talk given at the 2nd International Conference 
on $B$ Physics and CP Violation, Honolulu, Hawaii, March 1997; 
F.~W\"urthwein, hep-ex/9706010; J. Smith, these proceedings.
\bibitem{loso}D. London and A. Soni, {\it Phys.\ Lett.}~{\bf B407}, 61 (1997).
\bibitem{PAPIII}R. Fleischer, {\it Phys.\ Lett.}~{\bf B365}, 399 (1996).
\bibitem{PAPI}A.J. Buras and R. Fleischer,  {\it Phys.\ Lett.}~{\bf B365}, 390
(1996).
\bibitem{fm2}R. Fleischer and T. Mannel, hep-ph/9704423.
\bibitem{ag}A. Ali and C. Greub, hep-ph/9707251.
\bibitem{fm3}R. Fleischer and T. Mannel, hep-ph/9706261.
\bibitem{gnps}Y. Grossman, Y. Nir, S. Plaszczynski and M. Schune, 
hep-ph/9709288.

\end{thebibliography}
\end{document}